\documentclass[aps, prb, amsmath,amssymb, reprint, showkeys, longbibliography, superscriptaddress]{revtex4-2}

\usepackage[english]{babel}
\usepackage{graphicx}
\usepackage{dcolumn}
\usepackage{bm}
\usepackage{textcomp}
\usepackage[utf8]{inputenc}
\usepackage[T1]{fontenc}
\usepackage{mathptmx}

\usepackage{listings}
\usepackage{xspace}
\usepackage{enumitem}

\newcommand{\ie}{\emph{i.e.}\@\xspace}
\newcommand{\eg}{\emph{e.g.}\@\xspace}

\makeatletter
\def\fnum@figure{\textbf{Figure~\thefigure}}
\def\fnum@table{\textbf{Table~\thetable}}
\makeatother

\begin{document}

\setlength{\abovedisplayskip}{4pt}
\setlength{\belowdisplayskip}{4pt}

\title{Declarative Parameter Management for Automated Beamline Data Analysis: The Spice Concept}

\author{F.~L.~Boariu}
\affiliation{Helmholtz Zentrum Berlin, Berlin, Germany}
\author{M.~R\"ossle}
\affiliation{Helmholtz Zentrum Berlin, Berlin, Germany}
\author{A.~von~Reppert}
\affiliation{Institut f\"ur Physik und Astronomie, Universit\"at Potsdam, Potsdam, Germany}
\author{M.~Bargheer}
\affiliation{Helmholtz Zentrum Berlin, Berlin, Germany}
\affiliation{Institut f\"ur Physik und Astronomie, Universit\"at Potsdam, Potsdam, Germany}

\date{\today}

\begin{abstract}
  Automated data analysis at synchrotron beamlines is often limited by the
  complexity of managing many processing parameters across large sets of related
  measurements. Although the conversion of raw detector data into physically
  meaningful representations is largely deterministic, existing approaches
  typically embed parameter handling directly in analysis scripts, making
  automation, reusability, and reproducibility difficult. We introduce the
  spice
  concept, a declarative approach for managing analysis parameters in a
  scan-aware manner. Spice separates parameter definitions from the analysis
  code and organises their validity along the chronological sequence of scans
  using explicit update boundaries. This allows parameters to be applied
  consistently across multiple scans while still supporting local adjustments
  where necessary. Using X-ray diffraction as an example, we demonstrate how
  spice enables a clear separation between routine, instrument-specific data
  preparation (``cooking'') and the subsequent scientific interpretation. We
  describe an implementation within the Nx5d framework and discuss how the
  approach supports reproducible analysis, automated reprocessing, and
  provenance tracking. The spice concept provides a general architectural
  pattern for scalable and maintainable data processing at scientific
  installations.
\end{abstract}

\maketitle

\section{Introduction}
Scientific research infrastructures, synchrotrons and laboratories alike,
generate
experimental datasets whose volume and complexity render manual, scan-by-scan
analysis impractical. In response, numerous general-purpose workflow frameworks
have been developed to automate reduction pipelines and provide integrated
analysis environments
\cite{basham2015dawn,denolf2024ewoks,arnold2014mantid,liu2022mamba}. Complementing
these efforts, domain-focused platforms exist for many technique families; in
diffraction, for example, systems such as Pydidas\cite{storm2025pydidas},
DPDAK\cite{benecke2014dpdak}, and StreamSAXS\cite{wang2024streamsaxs} provide
end-user pipelines tailored to specific classes of experiments.

A central task in many experiments is the transformation of raw detector signals
into physically meaningful representations.
While these transformations are
typically deterministic and implemented in well-tested, community-driven
libraries (\eg \textit{pyFAI}, \textit{xrayutilities}, and \textit{silx})
\cite{ashiotis2015pyfai,kriegner2013xrayutilities,silx_website}, their execution
depends on numerous instrument- and experiment-specific parameters that change
during a beamtime. In practice, this knowledge is often encoded implicitly in
scripts and notebooks or distributed across facility software stacks and file
formats. As a result, automation, provenance capture, and large-scale
reprocessing are challenging. Achieving the FAIR principles for reusable open
data \cite{wilkinson2016fair} therefore requires preserving not only raw data
and algorithms but also the history of the parameters that render
the computation reproducible.

Despite the rich software ecosystem, one aspect of
ex\-pe\-ri\-men\-tal knowledge often remains insufficiently formalized: the explicit
representation of how parameter states evolve across sequences of
measurements. Existing workflow engines describe computational dependencies,
domain pipelines implement reduction strategies, and data-management platforms
ensure preservation and distribution. However, the res\-pon\-si\-bi\-li\-ty for
maintaining consistent pa\-ra\-me\-ter histories frequently falls back to ad hoc
conventions, personal expertise, or manual bookkeeping. As experiments increase
in scale and complexity, this implicit handling becomes a limiting factor for
reproducibility, portability, and reliable reprocessing. This motivates
a mechanism that treats parameter evolution itself as a first-class,
machine-readable entity, independent of individual scripts or software
environments, yet capable of interfacing with them.

In this work, we introduce the \emph{spice} concept, an architectural pattern
for declarative, scan-aware parameter management at the proposal level.
Its contribution is not merely that of another workflow engine or processing
library, but of a state model for parameter evolution for an ordered sequence
of scans.
Spice separates parameter definitions from analysis code, records explicit update
boundaries along the chronological scan sequence, and materializes per-scan
parameter views on demand. This provides stable defaults while allowing
well-defined local overrides. Using X-ray diffraction as a case study, we show
how spice enables scalable “cooking” of raw data into reusable, device-agnostic
representations, strengthen provenance, and support automated reprocessing and
second-pass analyses.

We then discuss how a proposal-level spice layer relates to the established
scan-oriented NeXus data format
\cite{konnecke2015nexus,nexusformat_website}.

Finally, we present Nx5d as one
concrete implementation that demonstrates the advantages;
however, the focus of this article is on the
architectural principles and their practical implications rather than on a
particular software realization.

\section{Representative Beamline Data: \\Scans, Frames, and Scan Types}

\begin{figure}[tbp!]
  \includegraphics[width=\columnwidth]{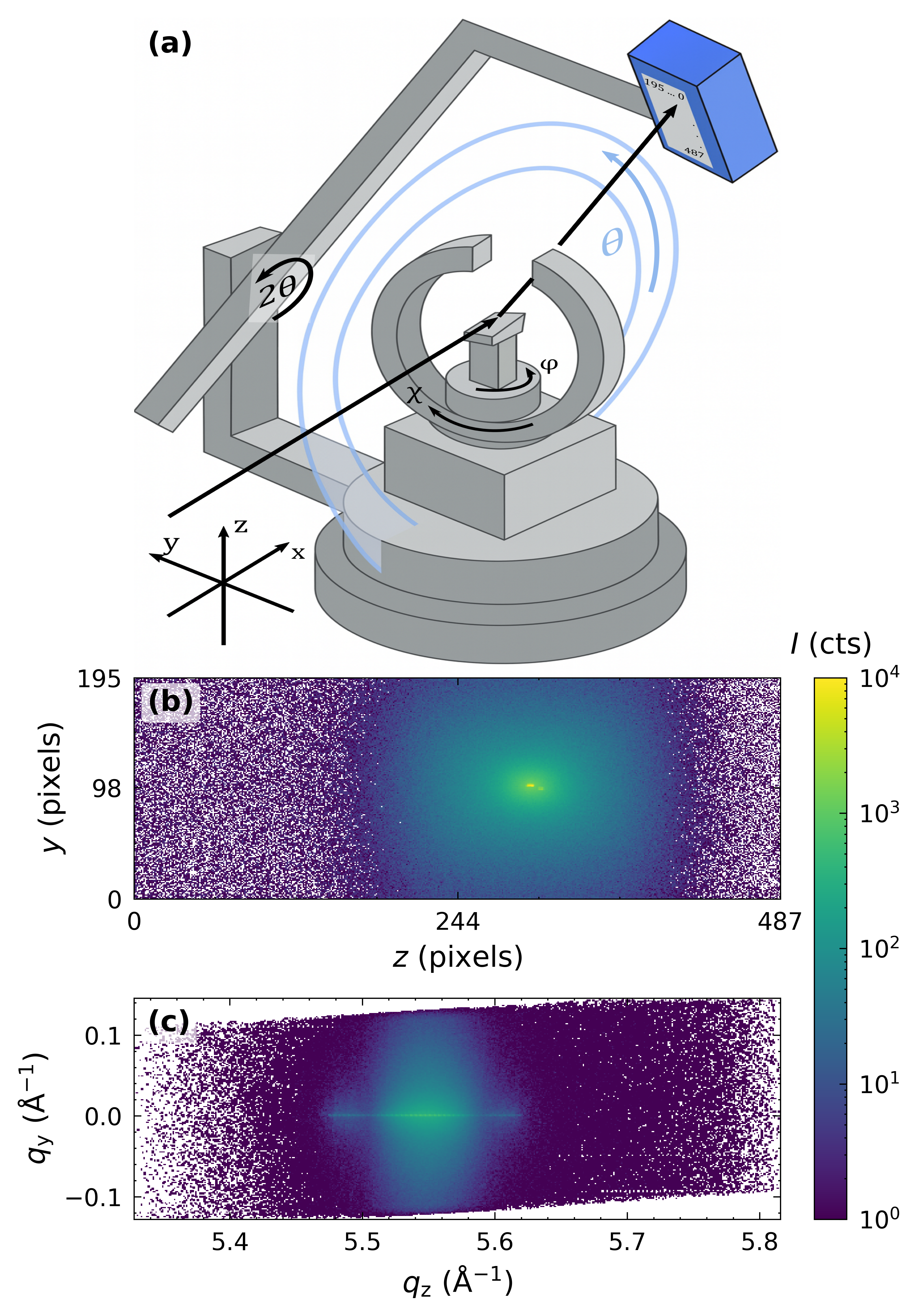}
  \caption{ (a) Schematic of the XRD experiment (adapted from
    \protect\cite{kriegner2013xrayutilities}).
    A crystalline sample is mounted on a
    multi-axis goniometer and illuminated with a X-ray beam, while the
    diffracted intensity is recorded on a 2D detector. (b) Example of raw data
    obtained by summing 31 detector images, yielding a reciprocal-space map of
    the Pt (222) reflection at $E = 6.8\,\mathrm{keV}$ and an incidence angle of
    $\approx 55^{\circ}$, displayed in pixel coordinates. (c) The same dataset
    transformed into reciprocal-space coordinates, providing a physically
    meaningful representation of the diffraction pattern from which quantitative
    analysis and physical interpretation can proceed.
    \label{fig:gonio}
  }
\end{figure}

To illustrate the data model and the application of the spice concept, we employ
data from an X-ray diffraction (XRD) experiment as a representative experimental
case study, reflecting our primary area of expertise.  A schematic
representation of a typical XRD experiment is shown in
Fig. \ref{fig:gonio}(a). In this setup, a sample is oriented in
space using a multi-axis goniometer, and the resulting diffraction pattern is
recorded by an area detector.  The acquired dataset comprises the angular
coordinates associated with each measurement position, along with
additional experimental parameters such as sample temperature, time
delays between X-ray and other context-defining variables.

The raw diffraction intensity data, exemplarily shown in
panel (b), have very limited direct scientific utility. Only after the
computation of the reciprocal space map (RSM), Fig.~\ref{fig:gonio}(c), which
represents the same intensity values on a grid in reciprocal ($Q$) space, can a
meaningful analysis be conducted.

A typical week of beamtime comprises several hundred individual
\emph{scans}. The scan constitutes the fundamental unit of an XRD experiment,
representing the smallest self-contained entity that can be subjected to
quantitative analysis and scientific interpretation. Each scan consists of a set
of datasets, \ie, N-dimensional arrays, that contain measurement and/or
positioner values recorded along a well-defined trajectory through the relevant
parameter space.

\begin{figure*}[t]
\centering
\begin{minipage}{\textwidth}
  \includegraphics[width=\columnwidth]{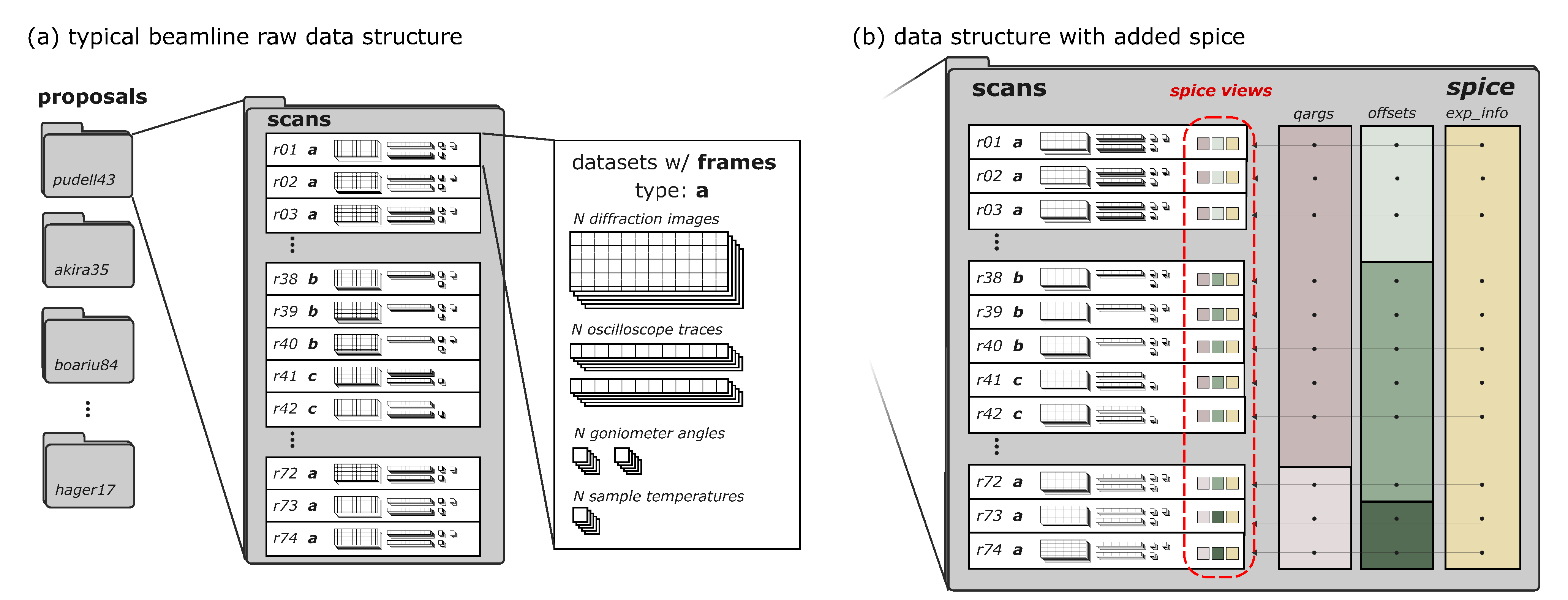}
\caption{ Panel (a): Data repository model of an X-ray diffraction
  facility. Independent scientific topics (``proposals'', left) are organized in
  separate folders or organizational units. Within each proposal, potentially
  heterogeneous series of measurements (``scans'', bold types \textbf{a},
  \textbf{b}, \dots, center) are stored and labelled chronologically (``run
  01'', ``run 02'', \dots). Each scan contains an arbitrary number of
  $N$-dimensional datasets, each typically sharing the same number of entries
  (``frames'', right) along its first dimension.  Panel (b): Extended data model
  for spice-enabled experiments. The spice database is a separate
  construct, maintained in parallel to the scans database. Spice update
  boundaries (anchors), shown as black horizontal lines between regions of
  different shading, reference scans by name. Dynamic spice blends (views) are
  generated on demand for each scan.
\label{fig:data-model}
}
\end{minipage}
\end{figure*}

Scans are heterogeneous: two given scans may comprise different kinds of
datasets or contain different numbers of frames \(N\). Nevertheless, the
hundreds of scans acquired in a typical XRD experiment usually form clusters of
a few dozen scans each, with each cluster consisting of structurally identical
scans. Consequently, the experiment as a whole comprises only a small (typically
single-digit) number of distinct scan categories. These include: alignment scans
(involving either translational displacements along the Cartesian coordinate
axes or rotational motions about the X-ray beam axis), reciprocal space mapping
(RSM) scans under static conditions (\ie without dependence on external
parameters), and parametrized reciprocal space maps \cite{schick2013ultrafast} or
reciprocal space slices \cite{zeuschner2021rss} (\ie, datasets acquired under
systematic variation of external control parameters such as sample temperature,
laser excitation conditions, pump–probe delay, applied voltage, or magnetic
field).

This variability is generally acknowledged implicitly by experimental
scientists, as reflected, for example, in the si\-mi\-la\-ri\-ty of the data-processing
algorithms they select for scans of the same kind. In contrast, we explicitly
introduce the concept of a \emph{scan type}. It is recorded as metadata for
each scan and captures the intent of the scientist at the moment the scan
is started in a codified manner.
This is illustrated in Fig.~\ref{fig:data-model}(a) by the small bold
letters {\textbf a}, {\textbf b}, and {\textbf c} adjacent to the scan
identifiers or \emph{keys} (\eg. ``r01'', ``r02'', etc.).

The resulting variational structure of scans constitutes the single most crucial
feature enabling a high degree of automation in the analysis workflow. The scan
type serves as the central attribute by which we associate each scan with the
appropriate, type-specific processing algorithms. Recognizing and formalizing
these recurring scan categories is essential for scalable, automated data
processing.

\section{Decoupling Data Pre-Processing \\ from Scientific Evaluation}

\begin{figure}[tbp!]
  \includegraphics[width=1.0\columnwidth]{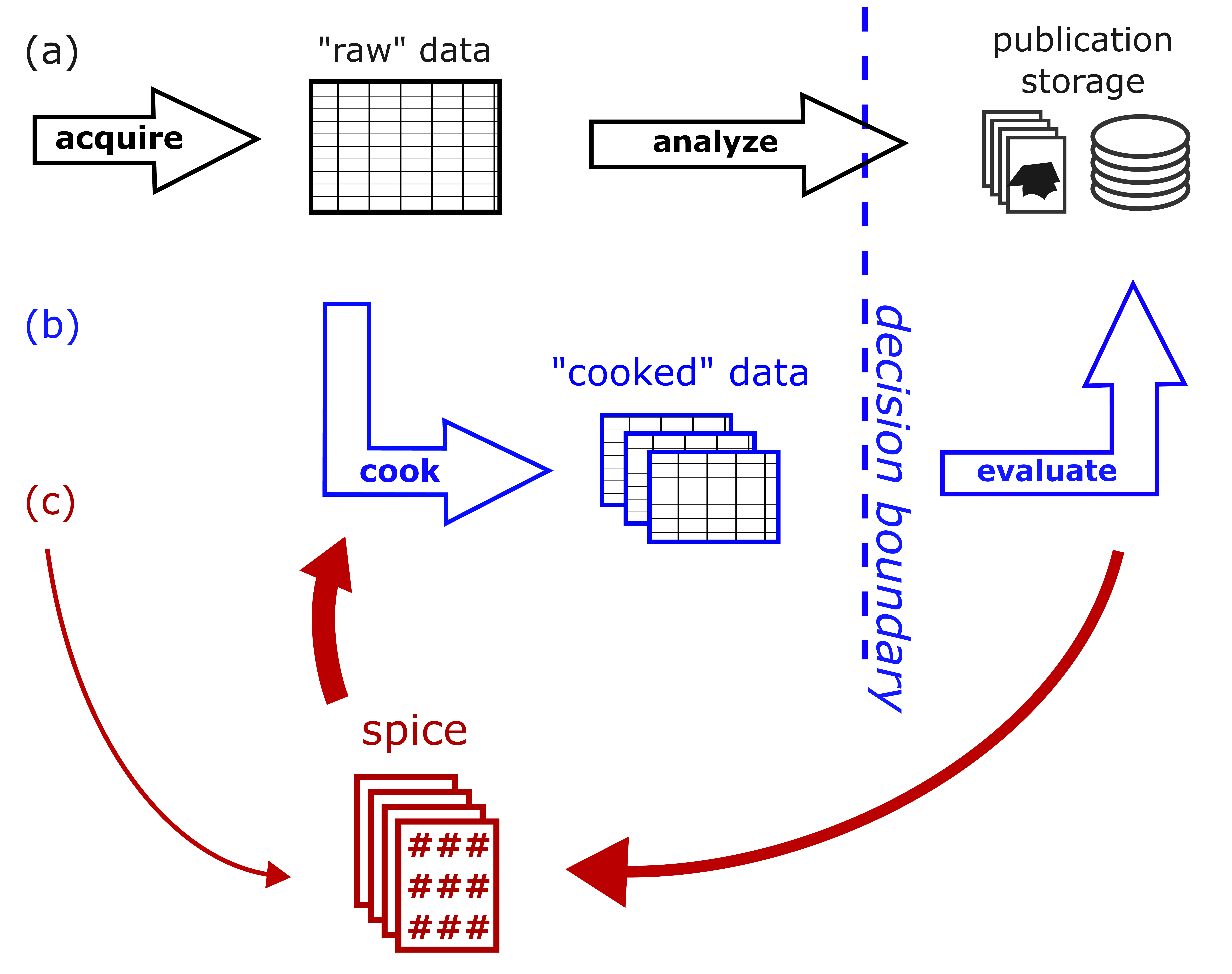}

\caption{Workflow models and information flow. (a) Black pictograms illustrate
  the traditional linear ``ac\-quire--anal\-yse--pro\-cess'' model, whose central
  artefact is the \emph{raw-data} stack. (b) Blue pictograms depict a
  differentiated model that introduces an additional \emph{cooked-data}
  artefact, produced by a device-dependent cooking step and subsequently used in
  an evaluation step that is largely device-independent and instead driven by
  the scientific question. The generation of \emph{cooked-data} requires no
  human decision-making, shifting the ``decision boundary'' into the evaluation
  step. (c) Red pictograms show how information required for cooking can flow
  back into the workflow (red arrows), yielding a cyclic process.
  The \emph{spice} artefact (red rectangles)
  provides initial default cooking parameters at first,
  and is subsequently modified by later updates.}
\label{fig:workflow}
\end{figure}

The conventional data-processing model is schematically illustrated in
Fig.~\ref{fig:workflow}(a).  The ``ac\-quire–anal\-yze–pro\-cess'' sequence,
represented by the black pictograms, reflects the typical conceptual framework
underlying a generic approach: immutable \emph{raw} data, initially generated by the
measurement apparatus, are subjected to user-driven analysis; upon completion of
this analysis, a distilled form of information is stored or disseminated in a
persistent format, for example as a peer-reviewed publication, a conference
presentation, or a static document such as a PDF containing final plots.

In contrast, we propose a processing model that decomposes the monolithic
``analysis'' stage into two distinct phases, ``cook'' and ``evaluate'', as
indicated by the red pictograms in Fig.~\ref{fig:workflow}(b).

``Cooking'' denotes the transformation of raw data into \emph{cooked data}, \ie,
into a physical, device-independent coordinate representation. Such a
representation constitutes, directly or indirectly, the basis for any subsequent
scientific evaluation. For a given experimental apparatus and scan type, the
cooking procedure is fixed and independent of the specific scientific question
being addressed. It does not intrinsically require human decision-making:
it may necessitate (possibly human-guided) adjustment or calibration of
its parameters, but the extent of intervention is primarily limited to
situational judgement of parameter fitness for a fixed purpose.
No inhrerently human selection based on scientific intent is necessary at
this point.

``Evaluation'', by contrast, encapsulates the core scientific inquiry and,
unlike cooking, is fundamentally driven by decision-making. Crucially, it
operates independently of the engineering-specific representation of the data
and is formulated solely in terms of the cooked data.

Up to this point, we have established a framework conducive to automation by
partitioning the workflow along a decision-making boundary, but this alone does
not resolve the automation challenge; it merely relocates it. A substantial
decision-making component remains in the form of specifying potentially hundreds
of parameter variations required to drive the cooking algorithm. To elucidate
why this constitutes a bottleneck, we consider an example analysis of the data
illustrated in Fig.~\ref{fig:data-model}(a).

In the most ideal case, the researcher has optimized their raw-data-processing
(“cooking”) procedures into concise one-line functions that accept only the raw
input data and a set of processing parameters as arguments. For example:

\begin{lstlisting}[
  language=Python,
  breaklines=true,
  breakatwhitespace=true,
  columns=fullflexible,
  basicstyle=\ttfamily\small
]
def cook_type_a(input_data, **params): ...
def cook_type_b(input_data, **params): ...
def cook_type_c(input_data, **params): ...
\end{lstlisting}

Furthermore, assume that the parametrization of the experiment’s processing is
captured by three objects:
\texttt{exp\_info} (describing the experimental geometry),
\texttt{offsets} (capturing angular offsets), and
\texttt{qargs} (providing configuration for the
transformation into reciprocal space, Q-space).
The selection of parameters here is just for illustration,
a real-life experiment would possibly require a different, or
more extensive, set of parameters.

Specifically, these objects can be represented as JSON-like
structures, for example:
\begin{lstlisting}[
  language=Python,
  breaklines=true,
  breakatwhitespace=true,
  columns=fullflexible,
  basicstyle=\ttfamily\small,
  keepspaces=true,
  showstringspaces=false
]
exp_info = { "goniometerAxes": [ ... ], ... }
offsets  = { "theta": 0, "phi": 0, "chi": 0, ... }
qargs    = { "dims": [ "qx", "qy" ], ... }
\end{lstlisting}

Under these conditions, a complete processing script for the experiment that
accounts for occasional changes in parameter values typically consists of
hundreds of lines that invoke the appropriate \texttt{cook\_*} functions with
varying parameter sets; for example:

\begin{lstlisting}[
  language=Python,
  breaklines=true,
  breakatwhitespace=true,
  columns=fullflexible,
  basicstyle=\ttfamily\small,
  keepspaces=true,
  showstringspaces=false
]
# initial parameters
cook_type_a(input_r01, exp_info0, offsets0, qargs0)
cook_type_a(input_r02, exp_info0, offsets0, qargs0)
...
cook_type_a(input_r37, exp_info0, offsets0, qargs0)

# offsets changing and type changing from scan r38
cook_type_b(input_r38, exp_info0, offsets1, qargs0)
cook_type_b(input_r39, exp_info0, offsets1, qargs0)
cook_type_b(input_r40, exp_info0, offsets1, qargs0)

# scan type changing
cook_type_c(input_r41, exp_info0, offsets1, qargs0)
cook_type_c(input_r42, exp_info0, offsets1, qargs0)
...

# qargs and type changing
cook_type_a(input_r72, exp_info0, offsets1, qargs1)

# offsets changing yet again
cook_type_a(input_r73, exp_info0, offsets2, qargs1)
cook_type_a(input_r74, exp_info0, offsets2, qargs1)
...
\end{lstlisting}

The modifications applied to the \texttt{offsets} and \texttt{qargs} parameter
sets (\eg \texttt{offsets0}, \texttt{offsets1}, \texttt{offsets2}, \dots) thus
become mutually entangled with one another and with changes in the scan
type. This interdependence is illustrated by the coloured rectangles in
Fig.~\ref{fig:data-model}(b), as examined in greater detail below.

Documenting these parameters on a per-scan basis, as exemplified in the code
snippet above—for instance, in Jupyter notebooks or analysis scripts—is highly
restrictive for two principal reasons. First, it entails a substantial amount of
redundant effort: any modification of parameter values or their structure must
be propagated across many consecutive lines of code, even if the researcher has
carefully stored distinct parameter sets in separate, versioned variables.  In
principle, constructing iterative loops over subsets of scans should be
straightforward, given the repetitive structure of the data. In practice,
however, this is extremely challenging, because the validity boundaries between
parameter ``versions'' within any given parameter class rarely
coincide with the boundaries of another class, and may evolve over time.
Consequently, writing and
maintaining the ``second line of code'' for the analysis of a given scan type is
as complex as writing the first, although it could be substantially simpler.

Secondly, the parameter definitions are tightly coupled to a specific analysis
script or notebook and are therefore not readily reusable.

Any “second pass” at analysing the same data, for instance, when sharing the
dataset with a collaborator is essentially as labour-intensive as the original
analysis: reproducing part of the workflow in a different script requires
duplicating the parameter-handling logic, and reliably updating any parameter at
a later stage necessitates locating and modifying all corresponding instances.

As a result, not only does the analysis procedure but also the overall framework
for interpreting the data become inherently monolithic.  In contrast, a core
principle of current research practice is that researchers are expected to be
able to examine and reuse one another’s data with as little hindrance as
possible. This principle of enabling efficient “second passes” is
institutionalized in community mindsets such as FAIR data and Open Data.

The data artifact \emph{spice} in Fig.~\ref{fig:workflow}(c) introduces a
crucial structural abstraction: it persistently stores all parameters required
for cooking in a queryable form.
For an ordered sequence of scans $S = (s1, \dots, sn)$, each
spice parameter class defines a piecewise constant state along the sequence
axis.

The spice structure for the example experiment
is shown in Fig.~\ref{fig:data-model}(b).  The colored rectangles represent
successive versions of the \texttt{exp\_info}, \texttt{offsets}, and
\texttt{qargs} objects introduced above. The thick horizontal lines within the
coloured rectangles denote \emph{anchorts}: update boundaries in the spice
database, marking scans at which the parameter state may change.
Any modification of a spice value for a given type is guaranteed to propagate
within, and only within, the interval delimited by two consecutive anchors.
A spice view for a given scan is obtained by resolving the parameter state
valid at that scan position across all parameter classes.
In essence, spice can be:

\begin{itemize}[leftmargin=1.5em,itemsep=0pt,parsep=0pt,topsep=2pt,partopsep=0pt]
\item \emph{seeded}, \ie created and inserted into a proposal;
\item \emph{anchored}, \ie assigned an explicit update boundary;
\item \emph{updated}, \ie its key values are modified, with changes propagating
  only between two anchor points;
\item or \emph{viewed}, \ie queried from the context of a specific scan.
\end{itemize}

Beyond the intrinsic link to a scan sequence, spice operations are timeless:
they are to be understood exclusively with respect to designated anchor points,
not to any kind of intrinsic chronological order.
Consequently, for example, if a spice update affecting
\texttt{qargs} is introduced at any point in time for scan \texttt{r40}, the
updated value becomes effective for all scans located between the corresponding
anchor points, including scans that were acquired both prior to \texttt{r72} and
subsequent to \texttt{r00}.

By analogy, a modification of the \texttt{offsets} values is applicable to scans
\texttt{r38...r72}, and an update of \texttt{exp\_info} propagates across the
entire proposal, encompassing even those scans that have not yet been acquired,
such as a possibly upcoming scan \texttt{r75}.
In operational terms, this semantics implies that data
processing systems must identify any cooked data products generated from a
superseeded spice version as obsolete and reprocess all data affected by the
corresponding spice updates.

Spice therefore constitutes a mechanism for specifying fine-grained
configuration parameters at the level of in\-di\-vi\-du\-al scans where required, while
otherwise defaulting to meaningful and practically useful values in the absence
of explicit per-scan settings. To exploit the strongly clustered structure that
characterizes the experimental design, the spice database associated with a
given proposal must be maintained at the proposal level, even though its
configuration is ultimately instantiated and applied at the level of individual
scans.

In the following section, we discuss the practical consequences of this
architecture and demonstrate how declarative, scan-aware parametrization enables
more reusable processing code, improves provenance and documentation, and
supports robust automation through systematic reprocessing.

\section{Discussion of the benefits: Why spice?}

\subsection{Simultaneously streamlined data analysis and collaboration}

An important distinction of the spice database is that it promotes
\emph{frictionless} collaboration.
In addition to enabling a high degree of automation during the initial
analysis, no additional curation steps are required to
prepare data for sharing; the same structures that support day-to-day analysis
simultaneously render results portable and reproducible. Given the spice
repository, along with the data, it is straightforward to export the results
into explicitly scan-based data representations, enabling their publication in
standardized scan-based formats.
Data FAIRness does not resolve to an
additional task for the scientist, but rather a by-product of an already
necessary workflow. The ``R'' of FAIR---Reusability---becomes more than
an abstract criterion: it becomes
a practically usable property of the data.

In a representative beamtime \cite{pudell25} consisting of 464
scans across five scan types,
the processing state changed at only 19 distinct scan positions.
Spice therefore
allowed the effective processing configuration for all 464 scans to be
represented by a small number of state transitions rather than by explicitly
maintaining a separate parameter configuration for every scan.”

\subsection{Inbuilt experiment logging and data visualization}

A large fraction of an XRD experiment logbook is typically the scan list: a
chronological enumeration of measurements annotated with the scan type (\eg
\texttt{``RSM''}, \texttt{``RSS''}, \dots)
and a small set of type-specific parameters (\eg
$\theta$-range, number of images in a reciprocal space mapping (RSM)
or reciprocal space slicing (RSS) operation, \dots).

While scan lists may be generated dynamically, explicit scan typisation is often
neither required nor persistently stored. As a result, automatically recorded
parameters remain generic, and researchers maintain supplementary lists to
document measurement intent.

The \emph{spice} concept, with its intrinsic scan types, provides a natural
basis for automatically generating scan lists that are semantically linked to
the respective scan types.

Beyond scan parameters, logbooks also collect observations, caveats, plots, and
outcomes of preliminary analyses. Although this information is often
unstructured, it can be associated with structure elements of the experimental
data, a relationship exploited at least in tendency
by modern electronic logbook systems \cite{wakonig2024scilog}.

The inherent data model underlying the spice concept offers persistent
identifiers to tie configuration and annotations to. Spice itself then
provides well-defined anchor points for attaching log entries and strengthens
the linkage between primary data, user-supplied metadata, and provenance.

\subsection{Simplifying self-service architecture}

A recurring challenge in particular at large-scale facilities is timely, reliable
post-beamtime access to data. When substantial data processing is required, this
goes beyond copying raw files: users must be able to perform the necessary
transformations; however, many are not sufficiently familiar with
instrument-specific workflows.

Self-service kiosk systems address this by allowing users to download cooked
data, but their usefulness depends on either universally correct cooking or the
ability to apply small corrections. The latter scenario is precisely the use
case for which \texttt{spice} was designed.

Finally, spice facilitates caching strategies in such infrastructures. Since
cooked data are a deterministic function of raw data and \texttt{spice}, and
updates are explicit, cache invalidation and automated reprocessing become
straightforward operations.

\section{Relation between the spice concept and the NeXus data format}

NeXus is a common data format for neutron, X-ray, and muon experiments that
stores measurements along with metadata in a standardized, machine-readable
structure. The presented spice framework is largely agnostic with respect to the
physical storage layer. Its only fundamental requirements are that scan
identifiers are chronologically sortable. In this sense, spice can interoperate
with existing NeXus-based infrastructures. Moreover, both approaches share the
overarching objective of improving cooperation, portability, and reusability in
data analysis and visualization.

At the same time, the internal organization of spice information is, by design,
not a direct match to the prevailing NeXus abstraction boundaries. A key feature
of spice is value defaulting across scan ranges, which requires managing
configuration at the proposal level. NeXus, in contrast, organizes data
primarily around the \texttt{NXentry} group, which most closely corresponds to
an individual scan and therefore forms a natural unit for acquisition and
archival workflows. This introduces a structural incompatibility.

NeXus application definitions may contain components of the information required
to describe cooking parameterisations. If NeXus is used as the acquisition
format (\ie, at the start of the processing pipeline), one could attempt to
reconstruct spice values and, most importantly, anchors from recorded parameters
by detecting differences between successive \texttt{NXentry} instances. However,
such heuristics are intrinsically limited: recorded updates reflect the moment
at which a user decided to write a value, not necessarily the evolution of the
underlying experimental state. In practice, this distinction can be
operationally relevant and limits the extent to which application definitions
alone can substitute for an explicit pa\-ra\-me\-ter history. Moreover, such a
data-import step is intrinsically unidirectional, while spice updates are
of a cyclical nature.
For spice-enabled experiments this ultimately leaves the question of
``how to store data?'' unanswered by NeXus alone.

The advantages of spice suggest a complementary development path. If NeXus were
to support mechanisms for managing selected classes of parameters at the
proposal scope, while retaining its proven scan-centered organization for raw
data, it could combine the established definitions of NeXus with the state-aware
automation enabled by spice. Such an extension would make implicit operational
knowledge explicit, improve reproducibility, and reduce the need for heuristic
reconstruction during downstream processing.

At the downstream end of the pipeline, NeXus naturally aligns with spice-based
processing. Once datasets have been ``cooked'', they can be expressed in
standardized NeXus scan-based structures. Analysis software can then operate
within well-defined, application-specific coordinate systems, enabling the
exchange of algorithms across facilities and fostering collaborative
development. In this role, NeXus primarily serves as a final, portable exchange
format rather than as a vehicle for the complexities of early-stage data
transformation. Realizing the full potential of such a workflow would benefit
from increasingly precise and fine-grained application definitions. Beyond
standardizing representations of processed data, this may eventually include
formal descriptions of the associated cooking parametrizations, \ie spice-type
state models. However, pushing NeXus downstream then leaves a significant
data modeling gap at the acquisition phase---one that
is inadequately covered by NeXus itself owing to aforementioned
structural incompatibilities.
Conversely, optimizing architecture to prematurely close that gap,
\ie, by favoring immediate device-to-storage into NeXus alone,
inadvertently holds back conception of better workflows.

\section{Nx5d: an example implementation of the Spice concept}

The spice concept is independent of a particular software
implementation. As a practical example, we have implemented the concepts
described above in Nx5d, an open-source Python framework for
experiment-agnostic scan-oriented data analysis~\cite{nx5d_repo,nx5d_docs}.
The cooking process specific to X-ray diffraction is delegated
to procedures within the XRayDC package~\cite{xraydc_repo}, with
a signature similar to
\begin{lstlisting}[
language={},
breaklines=true,
breakatwhitespace=true,
columns=fullflexible,
basicstyle=\ttfamily\small
]
  proc(scan_type, input_data, **spice).
\end{lstlisting}
The mapping of scan types to procedures is performed using Nx5d's
central feature---its capability of providing scan-level settings as spice:
\begin{lstlisting}[
language={},
breaklines=true,
breakatwhitespace=true,
columns=fullflexible,
basicstyle=\ttfamily\small
]
  recipes = {
    "staticRSM": "pymod://xraydc.cooking#sRSM",
    "delayRSM": "pymod://xraydc.cooking#dRSM",
    ...
  }
\end{lstlisting}
The URL-compatible way of referencing procedures in the internal-use
spice type \texttt{recipes} is to ensure the system remains modularly
extensible to other kinds of processing pipelines.

Nx5d provides a concrete realization of the separation between experimental
data, spice, and the processing algorithms while deliberately keeping the
location and physical representation of the underlying data outside the
analysis code.

A central design element of Nx5d is the use of URL templates to address
repository contents. Rather than embedding filesystem paths or storage-specific
access mechanisms in an analysis program, a repository is described by a URL
containing placeholders for the proposal and scan identifiers. For example,
the same addressing scheme can describe data stored in a local file,
\begin{lstlisting}[
language={},
breaklines=true,
breakatwhitespace=true,
columns=fullflexible,
basicstyle=\ttfamily\small
]
  file:///data/{proposal}/data.h5#{scan}
\end{lstlisting}
or in an object store,
\begin{lstlisting}[
language={},
breaklines=true,
breakatwhitespace=true,
columns=fullflexible,
basicstyle=\ttfamily\small
]
  s3://repository/{proposal}/{proposal}.zarr#{scan}.
\end{lstlisting}

The repository abstraction resolves these templates and de\-le\-ga\-tes the actual
loading to the corresponding backend. Importantly, spice is addressed through
the same general mechanism. It is treated as data, but is maintained as a
separate repository because its proposal-level structure differs from that of
experimental data. Consequently, the analysis code does not need to distinguish
whether its input data or spice originate from a local filesystem, an S3
repository, or another supported backend~\cite{nx5d_docs}. The physical
storage location therefore becomes a property of the repository configuration
rather than of the scientific analysis.

This abstraction is also exposed through the \texttt{pm} command-line tool,
which provides a lightweight self-service interface with low administrative
overhead to Nx5d repositories.
For the public data of the XPP@KMC-3 beamline at BESSY-II (KMC3) \cite{roessle21},
here used for demonstration, the built-in \texttt{kmc3pub} repository bookmark
contains the connection information, URL conventions, and proposal/scan
naming rules. A user can therefore select a scan without knowing its
underlying S3 location. For example, after installation of a compatible
Nx5d version and the analysis algorithms:
\begin{lstlisting}[
    language={},
    breaklines=true,
    breakatwhitespace=true,
    columns=fullflexible,
    basicstyle=\ttfamily\small
  ]
  pip install "nx5d >=3.2, <3.3"
  pip install "xraydc >= 0.4, <0.5"
\end{lstlisting}
the public repository can be queried
for proposal \texttt{252-cw33-15328-pudell} and
scan \texttt{r0047}, with:
\begin{lstlisting}[
    language={bash},
    breaklines=true,
    breakatwhitespace=true,
    columns=fullflexible,
    basicstyle=\ttfamily\small
  ]
  pm -r kmc3pub               \
     -p 252-cw33-15328-pudell \
     -s r0047                 \
     scan export -f cooked -o r0047.zarr
\end{lstlisting}
The referenced proposal is the actual data
basis behind the X-ray diffraction study published in ref.~\cite{pudell25}.

The precise output format is selectable; \texttt{pm} can export raw or cooked
data in formats including HDF5, NetCDF, and Zarr. Thus, from the user's
perspective, the operation consists of selecting a scientific object and a
desired representation rather than locating files or assembling a processing
pipeline. The latter tasks are resolved by the repository and spice
configuration~\cite{nx5d_docs}.

The same scan can be accessed programmatically from an IPython session. For
example, the workflow can be addressed directly by using Nx5d URLs:

\begin{lstlisting}[
    language={python},
    breaklines=true,
    breakatwhitespace=true,
    columns=fullflexible,
    basicstyle=\ttfamily\small,
    keepspaces=true,
    showstringspaces=false
  ]

  import os
  os.environ.update({
    "NX5D_DATA_REPO": "s3://kmc3pub-data/{proposal}"
                      "/{proposal}.zarr#{scan}",
    "NX5D_SPICE_REPO": "s3://kmc3pub-spice/"
                       "{proposal}/spice",
    "AWS_DEFAULT_REGION":    "US",
    "AWS_ACCESS_KEY_ID":     "2YJHA4GN3TB57OILX5J0",
    "AWS_SECRET_ACCESS_KEY": "iM5e+0eATWher9KLsjhq"
                             "u7oq51fQJKw4HtSQyzGI",
    "AWS_ENDPOINT_URL": "https://s3.rz.uni-potsdam.de"
  })

  from nx5d import Repository
  repo = Repository()
  proposal = repo._252_cw33_15328_pudell
  scan = proposal.r0047

  raw = scan.raw
  spice_view = scan.spice
  cooked = scan.cooked
\end{lstlisting}

The first part of the code concerns itself with definition of
data and spice URLs as well as additional backend-specific
configuration variables.
The S3 credentials reproduced here are for read-only, public-use
of the KMC3 repository.
In the absence of specific initialization parameters, the
\texttt{Repository} class will fall back to these.
For a cleaner separation between data access and analysis, beyond
the scope of this brief demonstration, these would have been set up
by the local system administrator outside the code itself.

The second block of code identifies the proposal within the repository,
and the scan within the proposal, as class properties. TAB completion
and auto-generated tabular views are also available in environments
Jupyter \cite{jupyter_website} environments.

Finally, the three scan properties correspond directly to the conceptual
data products
introduced above: \texttt{raw} provides the measured data in the device-oriented
representation, \texttt{spice} provides the parameter view applicable to the
selected scan, and \texttt{cooked} provides the result of applying the
corresponding processing recipe. The important point is that the processing
code operates on the scan abstraction rather than on the underlying storage
representation. The same interface can consequently be used when the
repository is moved from an S3-backed dataset to a local copy, or when the
spice repository is supplied independently of the experimental data.
The ability to optionally decouple configuration from the analysis code is
just one logical ingredient of this paradigm.

Nx5d therefore illustrates the intended role of spice as an architectural
layer rather than as an analysis algorithm. The repository abstraction
decouples scientific identifiers from storage, while the spice abstraction
decouples the parameter state required for cooking from the implementation of
the cooking algorithms. Together they allow a scan to be treated as a
self-contained scientific object whose raw data, applicable parameter state,
and cooked representation can be accessed consistently by command-line tools
or analysis scripts.

\section{Conclusion}

At experimental facilities, automated data analysis pipelines are
frequently constrained less by the availability of suitable algorithms than by
the practical challenges of maintaining consistent parameterisations across
extended, heterogeneous scan sequences. In this work, we introduced the
spice concept as a declarative, scan-aware framework for parameter
management that renders these challenges explicit and systematically
addressable. By decoupling parameter definitions from analysis code, recording
update boundaries (anchors) along the chronological scan axis, and generating
per-scan parameter views on demand, spice facilitates robust default
configurations while supporting localized, well-documented parameter overrides.

Using X-ray diffraction as a representative application domain, we demonstrated
how spice sharpens the conceptual and operational separation between
instrument-specific, predominantly deterministic data preparation (“cooking”)
and subsequent scientific interpretation (“evaluation”). This separation
enhances the reusability of cooked data products, enables automated reprocessing
in response to updated calibrations, and strengthens data provenance by
elevating the complete history of parametrization to a first-class, queryable
artifact.

Beyond the specific implementation described here, spice is proposed as
an architectural pattern that can be generalized to other beamline modalities
and, more broadly, to experimental workflows in which parameters evolve along a
structured sequence of measurements.

Future work will focus on integrating laboratory log systems, advancing
validation methodologies and tooling to facilitate deployment,
and aligning more closely with community data standards to enable
seamless exchange of reproducible, second-pass analyses.

\bibliography{spice.bib}

\section*{Acknowledgments}
Measurements were performed using the \textsc{XPP@KMC-3} instrument at the
\textsc{BESSY II} electron storage ring operated by the Helmholtz-Zentrum Berlin
für Materialien und Energie (HZB).
We acknowledge fruitful discussions regarding
the final stages of the architecture with I.~Zyzak, S.~R.~Patel, S.~Kazarski
and W.~Smith at HZB,
W.~de~Nolf at the European Synchrotron Radiation Facility,
J.-E.~Pudell and J.~Wrigley at the European XFEL, and extended beta
testing efforts of S.~Zeuschner, M.~Herzog and F.~Baltrusch at the UDKM group
of University of Potsdam, as well as M.~Mattern at the Max-Born-Institut.
We acknowledge financial support by the Deutsche Forschungsgemeinschaft
(DFG, German Research Foundation) through CRC/SFB 1636, project ID 510943930,
project numbers A01 and Z03.

\section*{AI assistance statement}
Portions of the manuscript text were prepared with the assistance of a
generative language model to improve clarity and language. All content was
reviewed, edited, and approved by the authors, who take full responsibility for
the manuscript.

\section*{Author contributions}
FLB and MR conceived the project. FLB developed the Nx5d spice implementation,
and drafted the manuscript. AvR and MR provided methodological input, tested early
implementations, and contributed to the revision of the manuscript. MR operated
the XPP@KMC-3 instrument and contributed facility-specific expertise. MB
provided project supervision, resources, and funding acquisition. All authors
discussed the results and approved the final manuscript.

\end{document}